\pdfoutput=1

\documentclass[10pt,letter,twocolumn]{article}
\usepackage{f1000_styles}

\usepackage[super,comma,sort]{natbib}
\usepackage[colorlinks=true,urlcolor=blue]{hyperref}

\usepackage{amsmath}
\newcommand*{\tbi}[1]{{\textit{\textbf{#1}}}}

\newcount\BoxNum \BoxNum 1\relax
\makeatletter
\newcommand*{\boxlabel}[1]{%
  \protected@write \@auxout {}{\string \newlabel {box:#1}{{\the\BoxNum}}{}}%
  \advance\BoxNum 1\relax}
\makeatother

\newcommand*{\boldrule}{\hrule height 1.2pt}
\newcommand*{\noterule}{\medskip\boldrule\medskip}	% for notes

\def\citeyear{\citep}
\def\autocite{\citep}
\def\textcite{\citet}

\begin{document}

\title{Puzzles in modern biology.~III.~Two kinds of causality in age-related disease}
%\titlenote{blank}
\author[1]{Steven A.~Frank}
\affil[1]{Department of Ecology and Evolutionary Biology, University of California, Irvine, CA 92697--2525 USA, safrank@uci.edu}

\maketitle
\thispagestyle{fancy}

\parskip=4pt

\begin{abstract}

The two primary causal dimensions of age-related disease are rate and function. Change in rate of disease development shifts the age of onset. Change in physiological function provides necessary steps in disease progression. A causal factor may alter the rate of physiological change, but that causal factor itself may have no direct physiological role. Alternatively, a causal factor may provide a necessary physiological function, but that causal factor itself may not alter the rate of disease onset. The rate-function duality provides the basis for solving puzzles of age-related disease. Causal factors of cancer illustrate the duality between rate processes of discovery, such as somatic mutation, and necessary physiological functions, such as invasive penetration across tissue barriers. Examples from cancer suggest general principles of age-related disease.

\end{abstract}


\vskip0.5in
\noterule
Preprint of published version: Frank, S. A. 2016. Puzzles in modern biology.~III.~Two kinds of causality in age-related disease. F1000Research 5:2533, \href{http://dx.doi.org/10.12688/f1000research.9789.1}{doi:10.12688/f1000research.9789.1}. Published under a Creative Commons \href{https://creativecommons.org/licenses/by/4.0/}{CC BY 4.0} license.
\smallskip
\noterule

\clearpage

\subsection*{Introduction}

If you inherit certain mutations of the \textit{p53} gene, you have an increased risk of cancer\autocite{kamihara14germline}. If you do not inherit such mutations, but nonetheless develop cancer, your tumor likely has a somatically acquired mutation in the apoptotic pathways associated with \textit{p53}\autocite{vogelstein00surfing}. 

In each case, \textit{p53}-associated mutation has a causal effect on cancer.

The inherited mutation increases the \tbi{rate} of cancer development and shifts disease onset to earlier ages. Shift in age of onset defines a cause of cancer.

The physiological change, breakdown of apoptosis, provides a necessary \tbi{function} in cancer development. Physiological necessity defines a cause of cancer.

\subsection*{Duality of rate and function}

A factor that shifts the age of onset may not be important physiologically. 

For example, a rise in somatic mutation may increase the rate of breakdown in apoptosis. Rapid breakdown in apoptosis shifts the age of onset. In this case, increased mutation directly changes the rate of onset but does not itself directly change physiological function.

A factor that changes physiology may not shift the age of onset. 

For example, tumors often adapt their metabolism to hypoxic conditions\autocite{semenza12hypoxia-inducible,gilkes14hypoxia}. The necessary physiological changes may arise relatively rapidly in response to hypoxia. The functional changes are a necessary cause of tumor development. However, rapidly acquired changes do not causally influence the rate of cancer development or the age of onset.

The duality of rate and function recur. Each causal factor must be evaluated simultaneously in two dimensions. How does a causal factor alter the rate of tumor development? How does a causal factor alter the physiological function of the tumor?

\subsection*{Identifying causal factors}

What sort of evidence could we collect to show that a factor plays a causal role in cancer?

Shift in age of onset is often studied in experiments\autocite{frank07dynamics}. Start with a particular mouse genotype. Create a knockout variant that lacks expression of a particular gene. Compare the age of tumor onset between the initial and knockout types. If the incidence curve in the knockout shifts to earlier ages, then loss of the target gene is a potential cause of cancer.

In general, we can relate the change in a potential causal factor to the change in the rate of cancer development and age of onset. 

Alternatively, studies may focus on physiological function. Experimentally, one may reverse a physiological change and measure the abrogation of a cancerous state. Success points to a physiologically necessary function. 

In general, we can relate the change in a potential causal factor to the change in the physiological function of a tumor. 

Large datasets allow one to correlate changes with cancer. A strong correlation suggests a candidate cause. However, the correlation may identify a factor that either increases the rate of cancer development or has a necessary physiological function in tumors. 

\subsection*{Solving different puzzles}

Full analysis requires simultaneous study of rate and function. The relative roles of the two causal dimensions vary with particular puzzles.

\textit{Treatment} requires a dual focus on interfering with cancer's physiological function and on altering the rate of escape from treatment. One typically begins by finding a way to block an essential physiological function. An initially successful block loses value in proportion to the rate at which the tumor escapes control. 

\textit{Prevention} depends only on slowing the rate of onset. Physiologically important functions may provide targets for slowing onset. However, some processes may significantly slow the rate of onset yet be physiologically unimportant. For example, the rate of onset may be increased by wound healing associated with a temporary increase the rate of cell division, by increased epigenetic instability, or by increased mutagenesis. Reduction of these rate-enhancing processes aids prevention.

\textit{Early detection} may focus on direct evidence of functional change. Small precancerous tumors associate with cancerous changes in physiology. Elevated levels of specific markers associate with cancerous physiological changes. Alternatively, one may focus on indicators associated with rate processes that shift the age of onset. Such indicators suggest elevated risk and the need to screen more carefully for direct signs of physiological change.

\textit{Basic understanding of onset} ultimately depends only on rate. Each causal factor must be evaluated within the complex interacting ensemble of processes that determine the overall rate of onset\autocite{frank07dynamics}. One must study how change in a causal factor shifts the age of onset within a particular background of other rate processes. Although only rate matters, function provides clues about which factors may influence rate.

\textit{Basic understanding of physiology} depends only on function. An important function does not necessarily influence rate. 

\subsection*{Rate is the search, function is the find}

In general, the relation between rate and function is similar to the relation between the process of discovery and the actual discovery itself\autocite{frank16puzzlesII}. In tumor evolution, the duality becomes the relation between the processes that change physiological function and the physiological function itself. For example, somatic mutation and natural selection between cellular lineages are processes that change physiological function. Acquired ability to invade across tissue barriers is a common physiological function of tumors. 

\subsection*{Age-related disease}

Age-related disease expresses the same duality of rate and function. Factors that influence rate alter the timing of disease onset. Factors that influence physiological function may be important targets for treatment, prevention and early detection. 

Basic understanding always demands a clear separation of rate and function. Only from that two-dimensional perspective can one solve particular puzzles. The solutions inevitably express the interactions of rate and function.

\subsection*{Competing interests}
No competing interests were disclosed.

\subsection*{Grant information}
National Science Foundation grant DEB--1251035 supports my research.

{\small\bibliographystyle{unsrtnat}
\bibliography{age}}

\end{document}